\documentclass{aa}
\usepackage{graphicx}
\usepackage{txfonts}
\newcommand{\RSTAR}{\mbox{$R_{\star}$}}
\newcommand{\TEFF}{\mbox{$T_{\rm eff}$}}
\newcommand{\RSOL}{\mbox{$R_{\sun}$}}
\newcommand{\MSOL}{\mbox{$M_{\sun}$}}
\newcommand{\LSOL}{\mbox{$L_{\sun}$}}
\newcommand{\MBOL}{\mbox{$M_{\rm BOL}$}}
\newcommand{\MSOLPERYR}{\mbox{$M_{\sun}$~yr$^{-1}$}}
\newcommand{\micron}{\mbox{$\mu$m}}
\newcommand{\KMS}{\mbox{km s$^{-1}$}}
\newcommand{\HOH}{\mbox{H$_2$O}}
\newcommand{\PERSQCM}{\mbox{cm$^{-2}$}}
\newcommand{\PERCBCM}{\mbox{cm$^{-3}$}}
\newcommand{\HRK}{\mbox{HR\_K}}
\newcommand{\VMICRO}{\mbox{$\varv_{\rm micro}$}}
\newcommand{\MDOT}{\mbox{$\dot{M}$}}
\newcommand{\VEXP}{\mbox{$\varv_{\rm exp}$}}
\newcommand{\LOGG}{\mbox{$\log \varg$}}
\begin{document}
\title{
Spatially resolving the outer atmosphere of 
the 
M giant BK~Vir in the CO first overtone lines
with VLTI/AMBER
\thanks{
Based on AMBER observations made with the Very Large Telescope 
Interferometer of 
the European Southern Observatory. Program ID: 081.D-0233(A)
(AMBER Guaranteed Time Observation)}
}

\author{K.~Ohnaka\inst{1} 
\and
K.-H.~Hofmann\inst{1} 
\and
D.~Schertl\inst{1}
\and
G.~Weigelt\inst{1} 
\and
F.~Malbet\inst{2}
\and
F.~Massi\inst{3}
\and
A.~Meilland\inst{4}
\and
Ph.~Stee\inst{4}
}

\offprints{K.~Ohnaka}

\institute{
Max-Planck-Institut f\"{u}r Radioastronomie, 
Auf dem H\"{u}gel 69, 53121 Bonn, Germany\\
\email{kohnaka@mpifr.de}
\and
UJF-Grenoble 1 / CNRS-INSU, Institut de Plan\'etologie et d'Astrophysique de 
Grenoble (IPAG) UMR 5274, Grenoble, F-38041, France
\and
INAF-Osservatorio Astrofisico di Arcetri, Instituto Nazionale di 
Astrofisica, Largo E. Fermi 5, 50125 Firenze, Italy
\and
Observatoire de la Cote d'Azur, Departement FIZEAU, 
Boulevard de l'Observatoire, B.P. 4229, 06304 Nice Cedex 4, 
France
}

\date{Received / Accepted }

\abstract
{
The mass-loss mechanism in normal K--M giant stars with small variability 
amplitudes is not yet understood, 
although they are the majority among red giant stars.  
}
{
We present high-spatial and high-spectral resolution observations 
of the M7 giant BK~Vir with AMBER at the Very Large Telescope 
Interferometer (VLTI).  
Our aim is to probe the physical properties of the outer atmosphere 
by spatially resolving the star in the individual CO first overtone lines. 
}
{
BK~Vir was observed between 2.26 and 2.31~\micron\ 
using the 16-32-48~m telescope configuration with an angular resolution 
of 9.8~mas and a spectral resolution of 12000.  
}
{
The uniform-disk diameters observed in the CO first overtone lines are 
12--31\% larger than those measured in the continuum.  We also detected 
asymmetry in the CO line-forming region, which manifests itself as 
non-zero/non-$\pi$ differential and closure phases. 
The data taken 1.5 months apart show possible time variation 
on a spatial scale of 30~mas (corresponding to 3 $\times$ stellar diameter) 
at the CO band head. 
Comparison of the observed data with the MARCS photospheric model shows 
that whereas the observed CO line spectrum can be well reproduced by the 
model, the angular sizes observed in the CO lines are much larger than 
predicted by the model. 
Our model with two additional CO layers above the MARCS photosphere reproduces 
the observed spectrum and interferometric data in the CO lines 
simultaneously. 
This model suggests that the inner CO layer at $\sim$1.2~\RSTAR\ is very 
dense and warm with a CO column density of $\sim \!\! 10^{22}$~\PERSQCM\ and 
temperatures of 1900--2100~K, while the outer CO layer at 2.5--3.0~\RSTAR\ 
is characterized by column densities of $10^{19}$--$10^{20}$~\PERSQCM\ and 
temperatures of 1500--2100~K.  
}
{
Our AMBER observations of BK~Vir have spatially resolved the extended 
molecular outer atmosphere of a normal M giant in the individual 
CO lines for the first time.  
The temperatures derived for the CO layers are higher than or equal to the 
uppermost layer of the MARCS photospheric model, implying 
the operation of some heating mechanism in the outer atmosphere.  
This study also illustrates that testing photospheric models only with the 
spectra of strong molecular or atomic features can be misleading.  
}


\titlerunning{Spatially resolving 
the M giant BK Vir in the CO first overtone lines
}
\authorrunning{Ohnaka et al.}
\maketitle

\begin{table*}
\begin{center}
\caption {Summary of the AMBER observations of BK~Vir with the E0-G0-H0
  AT configuration. 
}
\vspace*{-2mm}

\begin{tabular}{l l c c r c c c c l}\hline
\# & Night  & $t_{\rm obs}$ & $B_{\rm p}$ & PA     & Seeing  & 
Airmass & DIT & Number     & Calibrator\\ 
   &        & (UTC)        & (m)        & (\degr)& (\arcsec)        &
        & (s) & of frames  & \\
\hline
1  & 2009 Jan 19 & 07:43:32 & 13.9/27.9/41.9 & 73 & 0.9 & 1.2 & 6 & 75 &
$\nu$ Hya$^{\dagger}$\\
2  & 2009 Feb 28 & 06:15:16 & 15.6/31.2/46.8 & 74 & 0.7 & 1.1 & 6 & 75 &
$\nu$ Hya$^{\dagger}$\\
3  & 2009 Apr 15 & 03:43:30 & 15.9/31.8/47.8 & 74 & 0.8 & 1.1 & 6 & 75 &
$\beta$ Crv\\
\hline
\multicolumn{10}{l}{Note: 
$^{\dagger}$ means that the calibrator was observed before and after the 
science target.  Seeing is in the visible.}

\label{obslog}
\vspace*{-7mm}

\end{tabular}
\end{center}
\end{table*}

\section{Introduction}
\label{sect_intro}

The mass loss in late evolutionary stages of low- and intermediate-mass 
stars ($1 \la M/\MSOL \la 8$), particularly in the red giant branch 
(RGB) and asymptotic giant branch (AGB), is important 
not only for the evolution of the star itself but also for the chemical 
enrichment of the interstellar medium.  
The mass loss in Mira-type stars in the AGB phase, which are characterized by 
the large-amplitude pulsation with a variability amplitude of $\Delta V 
\approx 6$~mag and a period of $\sim$1 year, has been observationally and 
theoretically intensively studied.  
However, normal (i.e., non-Mira-type) 
M giants show variability amplitudes of $\Delta V$ = 
1--2~mag, much smaller than those of Mira stars, without clear periodic variations. 
They are classified as semi-regular or irregular variables.  
Among the M stars listed in the General Catalog of Variable 
Stars (GCVS) 4.2 (Samus et al. \cite{samus09}), there are 1987 stars 
classified as semi-regular or irregular giants (denoted as SRA, SRB, SRD, 
LB, and L in the catalog), in contrast to 1524 stars 
classified as Mira variables (see, however, Lebzelter et 
al. \cite{lebzelter95} for the uncertainty in the classification in the 
GCVS).  There are also many semi-regular or irregular variables whose spectral 
type is unknown in the catalog.  
Therefore, normal M giants with small variability amplitudes 
may outnumber Mira stars considerably. 

Despite their small variability amplitudes and the absence of clear 
periodicities, normal M giants are experiencing mass loss with 
mass-loss rates comparable to those in (optically bright) Mira stars.  
Normal M giants in the AGB phase show 
mass-loss rates of $10^{-7}$--$10^{-6}$~\MSOLPERYR\ and expansion 
velocities of 5--15~\KMS\ (e.g., Gonzalez Delgado et al. 
\cite{gonzalez_delgado03}; Winters et al. \cite{winters03}; 
De Beck et al. \cite{debeck10}).  
There is also evidence of significant mass loss in RGB stars 
with mass-loss rates up to $10^{-6}$~\MSOLPERYR\ near 
the tip of the RGB (Origlia et al. \cite{origlia07}, \cite{origlia10}; 
Ita et al. \cite{ita07}).

The mass-loss rate (\MDOT) and expansion velocity (\VEXP) of 
our target, BK~Vir, are estimated to be $(1.5-4)\times10^{-7}$~\MSOLPERYR\ 
and 4--7.5~\KMS, respectively 
(Gonzalez Delgado et al. \cite{gonzalez_delgado03}; Winters et al. 
\cite{winters03}), which are approximately the same as in the 
prototypical Mira $o$~Cet with \MDOT\ = $(1-8)\times 10^{-7}$~\MSOLPERYR\ 
and \VEXP\ = 1.5--7~\KMS (Winters et al. \cite{winters03}).  
This illustrates the importance of the contribution of normal M giants 
to the chemical enrichment of the interstellar medium.  

For understanding the mass-loss mechanism in red giant stars, it is 
indispensable to probe the physical properties of the region between the 
upper photosphere and the innermost part of the circumstellar envelope, 
where the energy and momentum are expected to be deposited for the wind 
acceleration.  
It is now known from spectroscopy and high-spatial resolution observations 
in the IR that there is a dense, warm ($\sim$1000--2000~K) molecular outer 
atmosphere, the so-called MOLsphere, 
extending to a few \RSTAR\ not only in Mira stars but also in normal K--M 
giants with small variability amplitudes 
(e.g., Tsuji et al. \cite{tsuji97}; Tsuji \cite{tsuji01}; 
Perrin et al. \cite{perrin04}; Ohnaka \cite{ohnaka04}; 
Ohnaka et al. \cite{ohnaka05}; Takami et al. \cite{takami09}).  
This MOLsphere is considered to play an important role in 
driving mass outflows.  
In the case of Mira stars, the large-amplitude pulsation can fairly explain 
the presence of the extended atmosphere 
(e.g., Ohnaka et al. \cite{ohnaka06}; Wittkowski et al. \cite{wittkowski07}, 
\cite{wittkowski08}; Woodruff et al. \cite{woodruff09}). 
However, the amplitudes of the stellar pulsation are much smaller 
in normal K--M giants, and therefore, the origin of the MOLsphere 
in these stars is by no means clear.  
In addition to the MOLsphere, the detection of 
UV emission lines as well as the H$\alpha$ line suggests the presence of 
a chromosphere (e.g., Eaton \cite{eaton95}; 
Isabel P\'erez Mart\'inez et al. \cite{isabel_perez_martinez11} and 
references therein).  
This means that there are both hot and warm components in the outer 
atmosphere of normal K--M giants, and 
the spectral analyses of the CO lines in 
the UV and IR support this picture (e.g., Wiedemann et al. \cite{wiedemann94}; 
McMurry \& Jordan \cite{mcmurry00}).

The multi-component nature of the outer atmosphere is a key to understanding 
the mass-loss mechanism in red giants.  
IR spectro-interferometry, which enables spatially resolved spectroscopy 
of molecular or atomic spectral features with a spatial resolution much 
higher than achieved with single-dish telescopes, is effective for probing 
the physical properties of the outer atmosphere. 
While Mira stars have been intensively studied by IR spectro-interferometry 
as mentioned above, IR spectro-interferometric observations 
of normal red giants are still very scarce.  
Recently, Mart\'i-Vidal et al. (\cite{marti-vidal11}) have presented 
near-IR spectro-interferometric observations of the M6/7 giant 
RS~Cap in the CO first overtone bands and \HOH\ bands with a medium 
spectral resolution of 1500 using VLTI/AMBER.  
In this paper, we report on VLTI/AMBER observations of the M7 
giant BK~Vir in the individual CO first overtone lines near 2.3~\micron\ 
with a higher spectral resolution of 12000.  
BK~Vir shows no peculiar features and, therefore, is a good representative 
of normal M giants (its evolutionary status 
is discussed in Sect.~\ref{subsect_param}).  
It was selected also for technical reasons, such as the brightness and 
angular size appropriate for AMBER observations.  

The paper is organized as follows.  Our AMBER observations and data reduction 
are described in Sect.~\ref{sect_obs}.  The observational results are 
shown in Sect.~\ref{sect_res} followed by the modeling of the data 
presented in Sect.~\ref{sect_modeling}.  
The implications of the results of the modeling are discussed in 
Sect.~\ref{sect_discuss}.  
Conclusions are presented in Sect.~\ref{sect_concl}.

\section{Observations}
\label{sect_obs}

AMBER (Petrov et al. \cite{petrov07}) is the near-IR (1.3---2.4~\micron) 
spectro-interferometric instrument at VLTI.  It enables us to obtain 
information on the amplitude and phase of the complex Fourier transform 
of the object's intensity distribution with a 
spectral resolution of up to 12000 by combining three 8.2~m Unit Telescopes 
(UTs) or 1.8~m Auxiliary Telescopes (ATs). 
AMBER measures several quantities related to the complex Fourier transform of 
the object's intensity distribution: visibility, differential phase (DP), and 
closure phase (CP).  
The visibility (or also called visibility amplitude) contains information 
about the size and shape of the object.  
The DP can measure the deviation of the object's phase in a spectral 
feature from that in the continuum, and non-zero DP within a spectral feature 
means that the spectral feature-forming region is asymmetric.  
The CP is the sum of the measured phases around a closed triangle of baselines. 
The CP is always zero or $\pi$ for point-symmetric objects, and 
non-zero and non-$\pi$ CPs indicate an asymmetry in the object. 

BK~Vir was observed at 3 epochs between January and April 2009 with 
AMBER using the E0-G0-H0 telescope configuration with ATs, 
as summarized in Table~\ref{obslog}.  
We used the high-spectral resolution mode in the $K$ band (\HRK) 
with a spectral resolution of 12000 
in the wavelength range between 2.26 and 2.31~\micron, which includes 
the CO (2,0) band head at 2.294~\micron.  The VLTI fringe-tracker 
FINITO was used to stabilize the fringes for a Detector Integration Time 
(DIT) of 6~s.  This long DIT was necessary to achieve reasonable SNRs 
in the CO lines, where the correlated flux (i.e., flux $\times$ 
visibility amplitude) is very low. 

We reduced the data with amdlib 
ver2.2\footnote{Available at http://www.jmmc.fr/data\_processing\_amber.htm}. 
We took the best 80\% of the frames in terms of the fringe SNR 
(Tatulli et al. \cite{tatulli07}).  
The interferometric calibrators ($\nu$~Hya and $\beta$~Crv) were observed 
just before and/or after BK~Vir.  We adopted angular diameters of 3.40 and 
4.45~mas for $\nu$~Hya and $\beta$~Crv, respectively (Richichi \& Perschron 
\cite{richichi05}).  
The errors of the calibrated visibilities, DPs, and CPs were 
estimated as described in Ohnaka et al. (\cite{ohnaka09}).  

\begin{figure*}
\resizebox{\hsize}{!}{\rotatebox{-90}{\includegraphics{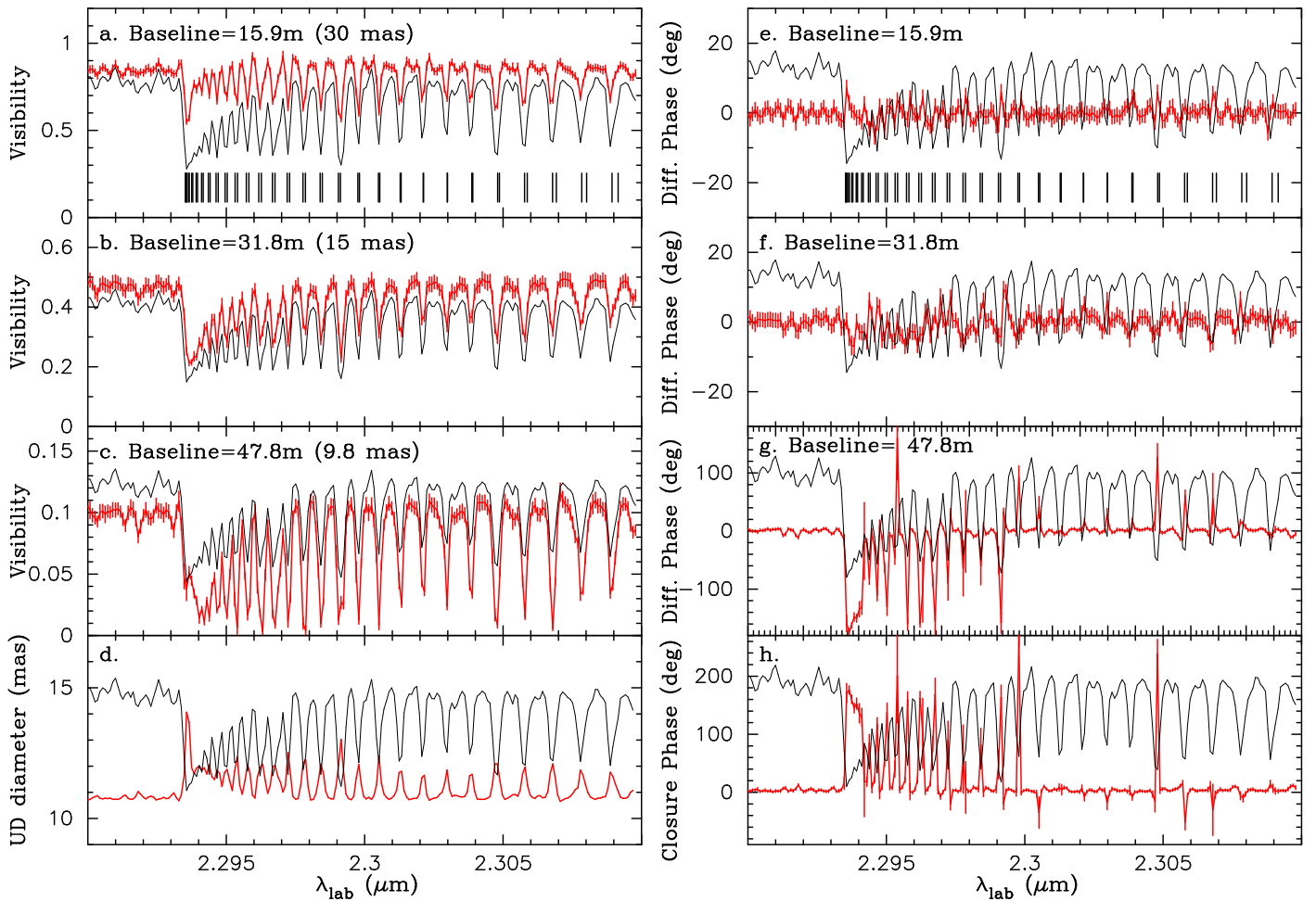}}}
\caption{AMBER data of BK~Vir (data set \#3).  
In each panel, the scaled observed spectrum is plotted by the black solid
lines.  
{\bf a})--{\bf c}) Visibilities observed on the 15.9, 31.8, and 47.8~m 
baselines (red lines).  The corresponding spatial resolutions are also given.  
{\bf d}) Uniform-disk diameter (red line) derived by fitting the visibilities 
shown in the panels {\bf a})--{\bf c}).  
{\bf e})--{\bf g}) Differential phases observed on the 15.9, 31.8, and 
47.8~m baselines (red lines). 
{\bf h}) Closure phase (red line). }
\label{obsres}
\end{figure*}

The AMBER data taken in \HRK\ mode before December 2009 are affected by 
the Fabry-Perot effect originating in the dichroics of the InfraRed 
Image Sensor (IRIS), which stabilizes the image motion 
\footnote{AMBER data taken after the 
replacement of the IRIS dichroics in December 2009 do not show this effect 
any more.}.  
This effect appears as time-dependent, high-frequency beating in raw 
visibilities, DPs, and CPs, when plotted as a function of 
wavelength.  Fortunately, this effect is mostly removed by using the 
calibrators observed just before and/or after the science target, 
although the residual of this effect appears as the small-amplitude 
beating in some cases (as seen in the observed CP at wavelengths below 
2.293~\micron\ in Fig.~\ref{obsres}e).

The long DIT of 6~s makes the absolute calibration of the observed 
visibilities uncertain, because we cannot know the fraction of time during 
DIT when the fringes were lost, and the fringe tracking may work differently 
for the science target and the calibrator.  
Therefore, we took the following approach for the absolute calibration of 
the observed visibilities.  
We note that the $K$-band uniform-disk diameters of BK~Vir reported in 
the literature show little time variation within the measurement errors: 
$11.2 \pm 0.6$~mas (Dyck et al. \cite{dyck98}), $10.73 \pm 0.23$~mas 
(Perrin et al. \cite{perrin98}), and $11.49 \pm 0.40$~mas (Perrin et al. 
\cite{perrin03}).  
In the present work, we adopted a uniform-disk diameter of 10.73~mas, 
which has the smallest error among the above measurements.  
We scaled the observed calibrated visibilities in the continuum to 
those expected from the uniform disk with an angular diameter of 10.73~mas.

The wavelength calibration was carried out using the telluric lines 
identified in the spectra of the calibrators.  
As a template of the telluric lines, we convolved the atmospheric transmission 
spectrum measured at the Kitt Peak National 
Observatory\footnote{http://www.eso.org/sci/facilities/paranal/instruments/isaac/tools/\\spectra/atmos\_S\_K.fits}
to match the spectral resolution of AMBER.  
The uncertainty in wavelength calibration is $4 \times 10^{-5}$~\micron\ 
($\sim$5.2~\KMS).  
The wavelength scale was converted to the laboratory frame using a 
heliocentric velocity of 16.5~\KMS\ measured for BK~Vir 
(Gontcharov \cite{gontcharov06}; Famaey et al. \cite{famaey05}). 

Since no spectroscopic standard star was observed, we used the interferometric 
calibrators for the spectroscopic calibration to remove the telluric lines 
from the observed spectrum of BK~Vir.  
However, both $\beta$~Crv (G5II) and $\nu$~Hya (K0/K1III) show CO absorption 
lines. 
In this case, the calibrated spectrum of the science target is 
derived as 
$F_{\star}^{\rm sci} = F_{\rm obs}^{\rm sci} / (F_{\rm obs}^{\rm
  cal}/F_{\star}^{\rm cal})$, 
where $F_{\star}^{\rm sci(cal)}$ and $F_{\rm obs}^{\rm sci(cal)}$ denote 
the true and observed (i.e., including the atmospheric transmission 
and the detector's response) spectra of the science target 
(or calibrator), respectively.  
To estimate the true spectrum of $\nu$~Hya, we used the high-resolution 
$K$-band spectrum of $\alpha$~Boo (K1III) obtained by Wallace \& Hinkle 
(\cite{wallace96}), which has a similar spectral type to $\nu$ Hya.  
The spectrum of $\alpha$~Boo, which has a spectral resolution of 89000, 
was convolved down to match the resolution of AMBER.  This convolved 
spectrum of $\alpha$~Boo reproduces the strength of the CO band 
head observed in $\nu$~Hya, 
indicating that it can be used as the true spectrum of $\nu$~Hya.

For $\beta$~Crv, however, no high-resolution CO spectra of stars with similar 
spectral types are available in the literature.  Therefore, 
we estimated the true spectrum for $\beta$~Crv by constructing a model CO 
spectrum using the MARCS models (Gustafsson et al. \cite{gustafsson08}). 
The MARCS models 
represent spherically symmetric photospheres with molecular and atomic 
line opacities taken into account using the opacity sampling method.  
Each MARCS model is specified by effective temperature (\TEFF), 
surface gravity ($\varg$), microturbulent velocity (\VMICRO), 
chemical composition, and stellar mass ($M_{\star}$).  
These parameters derived for $\beta$~Crv are  \TEFF\ = 5100--5145~K, 
$\log \varg$ = 2.2--2.7 (given in units of cm~s$^{-2}$ throughout the paper), 
\VMICRO\ = 1.5--2.1~\KMS, and 3.0--3.7~\MSOL\ (Luck et al. \cite{luck95}; 
Takeda et al. \cite{takeda08}; Lyubimkov et al. \cite{lyubimkov10}).  
The CNO abundances derived by Luck et al. (\cite{luck95}) suggest the mixing 
of CN-cycled material.  
We selected a model atmosphere with parameters as close to these 
values as possible from the MARCS model grid\footnote{http://marcs.astro.uu.se}. 
The selected model is characterized with \TEFF\ = 5250~K, $\log \varg$ = 2.5, 
$\VMICRO$ = 2~\KMS, $M_{\star}$ = 1~\MSOL, and ``moderately CN-cycled'' 
chemical composition with C/N = 1.5.  
The stellar mass of this model is smaller than the observationally 
estimated values above.  However, the atmosphere is approximately 
plane-parallel with a geometrical thickness of $\sim$1\% of the 
stellar radius.  In this case, the stellar mass does not affect the 
atmospheric structure noticeably.  
Using the temperature and pressure distributions of this model, we 
computed a synthetic spectrum for the CO first overtone lines 
using the line list of Goorvitch (\cite{goorvitch94}).  
The carbon abundance was adjusted to fit the CO band head in the 
uncalibrated spectrum of $\beta$~Crv.  
With $F_{\star}^{\rm cal}$ constructed in this fashion, 
the calibrated spectrum of BK~Vir was derived using the above 
equation.  
Although the strengths of the CO lines in two calibrators are 
very different (the line depths are 20--30\% and 10\% for $\nu$~Hya 
and $\beta$~Crv, respectively), the spectra of BK~Vir calibrated with 
these stars agree well (see Fig.~\ref{tempvar}a, where both spectra 
calibrated with $\nu$~Hya (2009 Feb 28) and $\beta$~Crv (2009 Apr 15) 
are shown).  This indicates that the calibrated CO line spectra of 
BK~Vir are not significantly affected by the above, unusual spectroscopic 
calibration.

\section{Results}
\label{sect_res}

Figure~\ref{obsres} shows the observed visibilities, DPs, CPs, and spectrum 
from the data set \#3.  
The results from the other data sets are similar, and therefore, not shown.  
The figure reveals significant signatures of the CO lines in the observed 
visibilities, DPs, and CPs.  
Thanks to the long DIT of 6~s, the visibilities as low as 0.01 could be 
measured in the CO lines. 
As plotted in Fig.~\ref{obsres}d, 
the uniform-disk diameter obtained from the observed visibilities shows 
that the star appears larger in the CO lines (12--14~mas) than in the 
continuum (10.73~mas) by 12--31\%.  
However, the uniform-disk fit in the CO lines is poor with 
reduced $\chi^2$ values much higher than 1 (as high as $10^2$--$10^3$), 
which means 
that the uniform disk is a poor representation of the object's shape 
in the CO lines.  Therefore, the uniform-disk diameter in the CO lines 
should be regarded as a very rough estimate of the characteristic size 
of the object.  

Figures~\ref{obsres}e--\ref{obsres}h show the detection of non-zero DPs 
(particularly clear on the 47.8~m baseline) and non-zero/non-$\pi$ 
CPs in the CO lines, which indicate an asymmetry in the CO line-forming 
region.   
High-spatial resolution observations have revealed that 
asymmetries in the atmosphere of red giant stars are common.  
The images of the Mira star R~Cas in the TiO bands obtained by 
Weigelt et al. (\cite{weigelt96}) show elongation, while the image in the 
continuum appears circular.  
The optical images of the prototypical Mira star $o$~Cet taken with 
the Hubble Space Telescope show significant asymmetries (Karovska 
et al. \cite{karovska97}).  
Tuthill et al. (\cite{tuthill99}) also revealed departures from 
circular symmetry in five Mira 
stars (including the above two stars) in the optical and near-IR.  
The aperture synthesis images of the Mira stars R~Aqr and U~Ori in the 
$H$ band obtained with the Infrared Optical Telescope Array (IOTA) show 
asymmetric, extended atmospheres (Ragland et al. \cite{ragland08}; 
Pluzhnik et al. \cite{pluzhnik09}).  
U~Ori shows asymmetry in the $K$ band as well (Mondal \& Chandrasekhar 
\cite{mondal04}).  
There are more stars for which asymmetries are indicated from the 
detection of non-zero CPs.  
Bedding et al. (\cite{bedding97}) detected non-zero CPs for the 
Mira-like star R~Dor.  
Ragland et al. (\cite{ragland06}) also report asymmetries 
in a sample of AGB stars based on the detection of non-zero CPs.  
More recently, Wittkowski et al. (\cite{wittkowski11}) have shown 
asymmetries in the CO first overtone bands in Mira stars 
by AMBER observations with a spectral resolution of 1500.  
Our AMBER observations are the first study to detect asymmetry in 
a red giant star in the individual CO lines.

\begin{figure}
\resizebox{\hsize}{!}{\rotatebox{0}{\includegraphics{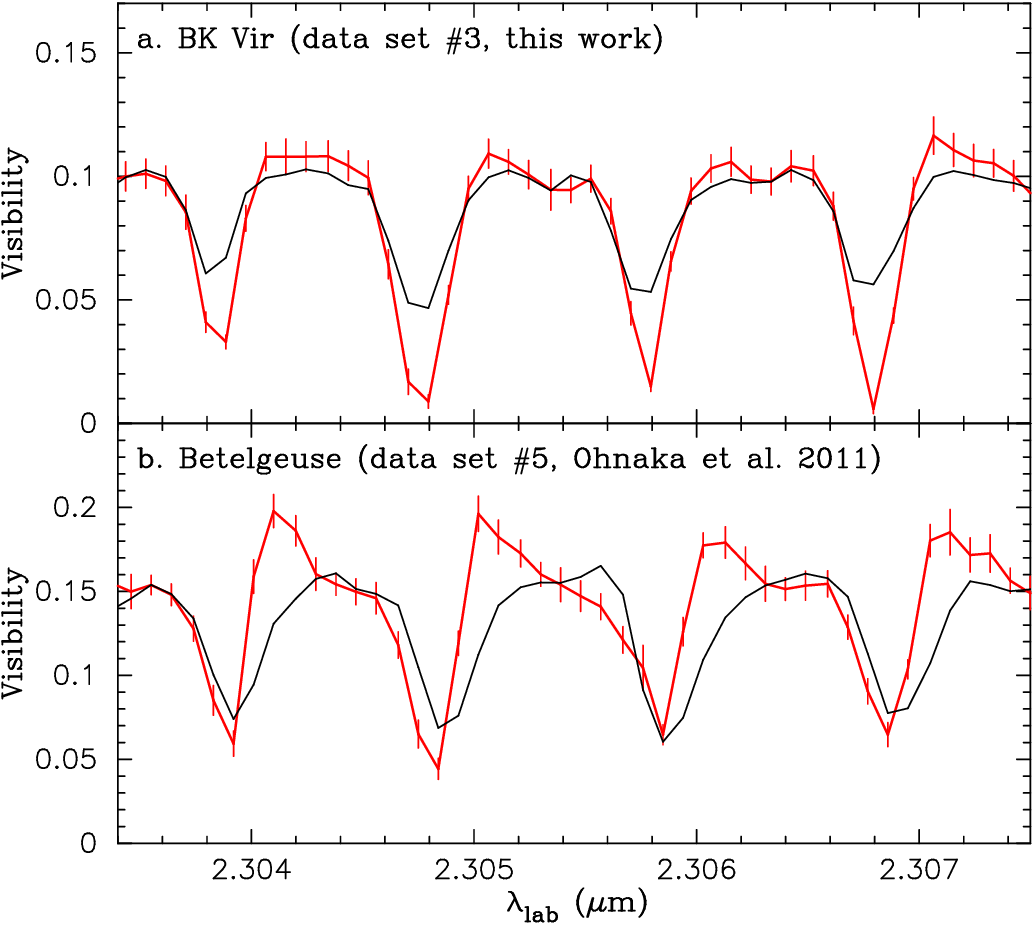}}}
\caption{Visibilities in four representative CO first overtone lines in the 
red giant BK~Vir (panel {\bf a}) and the red supergiant Betelgeuse 
(panel {\bf b}).  
In both panels, the black solid lines represent the scaled observed spectra. 
The red solid lines represent the visibility obtained for BK~Vir 
on the 47.8~m baseline (panel {\bf a}) and that obtained for Betelgeuse 
on the 11.54~m baseline (panel {\bf b}).  }
\label{obsresCO}
\end{figure}

Figure~\ref{obsresCO}a, which shows an enlarged view of the visibilities 
in four representative CO lines obtained on the 47.8~m baseline, 
reveals that 
the observed visibilities are approximately symmetric with respect to the 
line center. 
On the other hand, Fig.~\ref{obsresCO}b shows 
the visibility of the red supergiant Betelgeuse observed on the 11.54~m 
baseline  (data set \#5 of Ohnaka et al. \cite{ohnaka11}). 
Because the angular diameter of Betelgeuse (42.5~mas, Ohnaka et al. 
\cite{ohnaka11}) is four times larger than that of BK~Vir (10.73~mas), 
the 11.54~m baseline for Betelgeuse probes nearly the same 
spatial scale in terms of the stellar angular diameter 
as the 47.8~m baseline for BK~Vir (0.9 $\times$ stellar 
angular diameter).  
Figure~\ref{obsresCO}b clearly shows that the observed visibilities are 
asymmetric with respect to the line center in the case of Betelgeuse: 
the visibilities are characterized by the minima in the blue wing and 
the maxima in the red wing.  
These asymmetric visibilities are interpreted 
as temporally variable, inhomogeneous gas motions within 1.3--1.4~\RSTAR\ 
over the stellar surface 
with velocity amplitudes of up to 20--30~\KMS\ (Ohnaka et al. \cite{ohnaka09}, 
\cite{ohnaka11}).  
The absence of such asymmetric visibilities in 
BK~Vir suggests that the amplitude of the inhomogeneous velocity field is 
much smaller than the AMBER's spectral resolution of 12000, say, $\sim$5~\KMS, 
and/or the spatial scale of the inhomogeneous gas motions is much smaller 
than the spatial resolution of the current AMBER data (9.8~mas).  
However, the former possibility is more favorable, because 
the non-zero DPs and non-zero/non-$\pi$ CPs shown in Fig.~\ref{obsres} 
indicate that the inhomogeneity is spatially resolved with the current 
spatial resolution.  

\begin{figure}
\resizebox{\hsize}{!}{\rotatebox{0}{\includegraphics{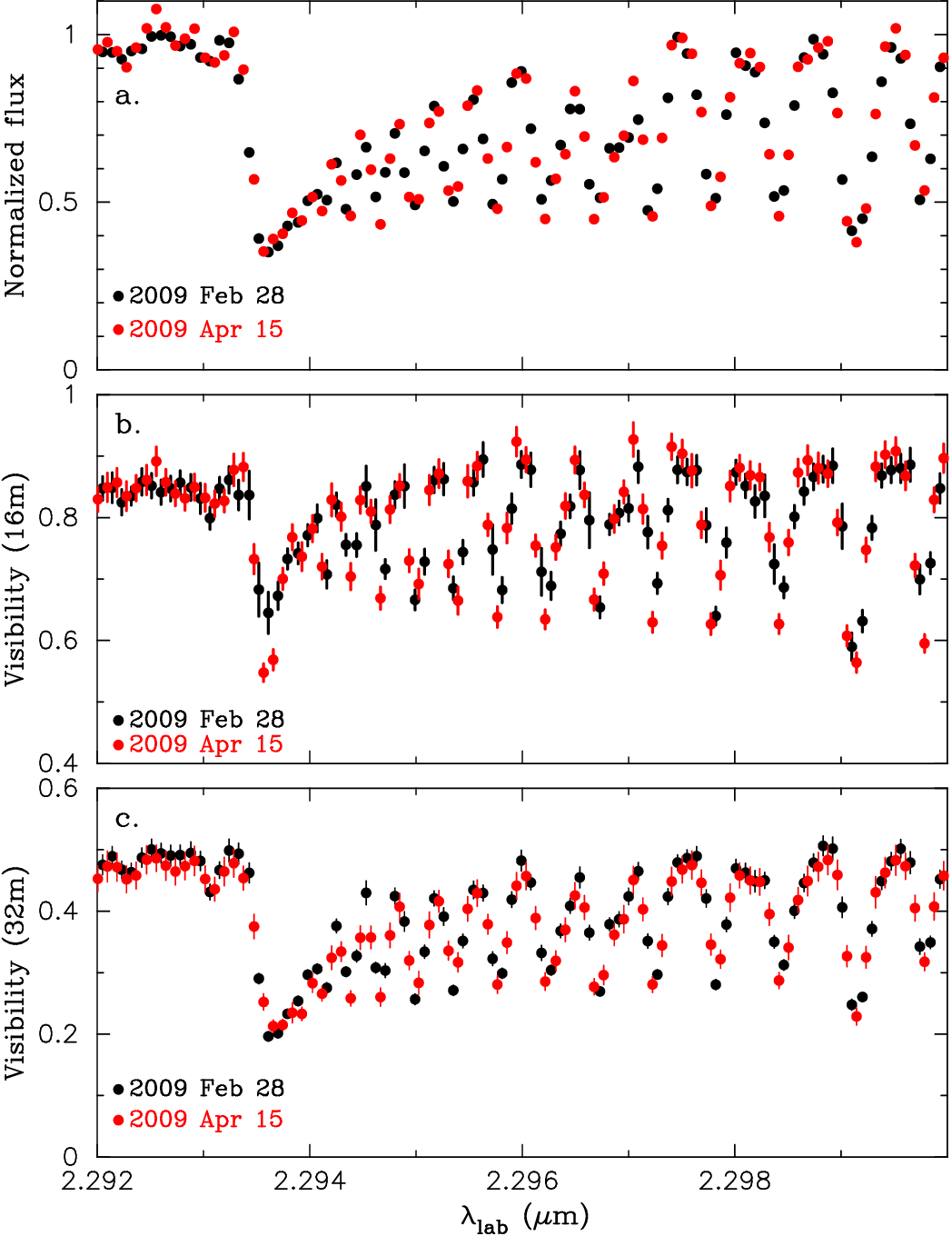}}}
\caption{Spectra (panel {\bf a}) and visibilities (panels {\bf b}: 16~m 
baseline and {\bf c}: 32~m baseline) obtained on 2009 Feb 28 and 2009 Apr 
15. 
In all panels, the data taken on 2009 Feb 28 and 2009 Apr 15 are plotted 
by the black and red dots, respectively. 
}
\label{tempvar}
\end{figure}

We examined time variation using the AMBER data taken at nearly the same 
$uv$ points.  
The shortest and middle baselines of the data sets taken on 2009 Feb 28 
and 2009 Apr 15 are almost the same (the 1~m difference in the longest 
baseline is too significant to study time variations).  
A comparison between these two data sets is shown in 
Fig.~\ref{tempvar}.  
The sampled wavelengths are slightly different for 
the two nights because of the difference in the correction to convert 
the observed wavelength scale to the heliocentric scale.  
Figures~\ref{tempvar}a and \ref{tempvar}c show that the observed spectra and 
the visibilities on the 32~m baseline 
measured on the two nights form continuous curves, suggesting no time 
variation.  This is also the case for the 16~m visibility shown in 
Fig.~\ref{tempvar}b except for the CO band head at 2.2936~\micron, 
where the visibility measured on Apr 15 is 12--20\% lower than that 
measured on Feb 28.  
This suggests possible time variation on a spatial scale of 30~mas 
(3 $\times$ stellar diameter).  
The decrease in the visibility corresponds to an increase in the 
uniform-disk diameter by 22\% (from 16 to 19.5~mas) or an increase in 
the Gaussian FWHM by 14\% (from 10.5 to 12~mas) between Feb 28 and 
Apr 15.  
However, the difference is within 3-$\sigma$ of the data from Feb 28.  
Therefore, more observations are needed to definitively confirm time 
variations in the visibility.

\section{Modeling of the AMBER data}
\label{sect_modeling}

\subsection{Determination of stellar parameters}
\label{subsect_param}

To interpret the AMBER observations, we use the MARCS models.  
As mentioned in Sect.~\ref{sect_obs}, these models assume spherical 
symmetry.  
Whereas the asymmetry in the CO-line-forming region is detected, 
we use these spherical models as the first approximation.  
To select a MARCS model appropriate for BK~Vir, we determined 
its stellar parameters (\TEFF, $\varg$, \VMICRO, $M_{\star}$, and chemical 
composition) as follows.  

We estimated the effective temperature from the observed angular 
diameter and the bolometric 
flux obtained by integrating the photometric data available in the 
literature 
(Two-Micron Sky Survey, Neugebauer \& Leighton \cite{neugebauer69}; 
NOMAD Catalog, Zacharias et al. \cite{zacharias04};
2MASS, Skrutskie et al. \cite{skrutskie06}; 
Fouque et al. \cite{fouque92}; 
Kerschbaum \& Hron \cite{kerschbaum94}; 
IRAS Point Source Catalog).  
The photometric data were corrected for the interstellar extinction 
using $E(B-V)$ = 0.023 (Schlegel et al. \cite{schlegel98}) and 
assuming $A_{V} = 3.1 E(B-V)$. 
Combining the angular diameter of $10.73$~mas adopted in 
Sect.~\ref{sect_obs} 
and a derived bolometric flux of $2.80 \times 10^{-9}$~\mbox{$\rm W m^{-2}$} 
results in an effective temperature of $2920$~K.  
While this agrees with the effective temperature of $3074 \pm 141$~K 
derived by Dyck et al. (\cite{dyck98}), they obtained a higher bolometric flux 
of $3.9 \times 10^{-9}$~\mbox{$\rm W m^{-2}$}.  
If we adopt, as the uncertainty in the bolometric flux, 
a half of the difference between the values from Dyck et al. (\cite{dyck98}) 
and from the present work, the total error (i.e., error resulting from 
the uncertainties in the angular diameter and in the bolometric flux) 
in our effective temperature is $\pm 150$~K.  
Using the distance of 180~pc based on the Hipparcos parallax of 
$5.53 \pm 0.68$~mas (van Leeuwen \cite{vanleeuwen07}) and the above 
bolometric flux of $2.80 \times 10^{-9}$~\mbox{$\rm W m^{-2}$}, we derived 
a luminosity of 2700~\LSOL\ (\MBOL\ = $-3.8$).  With the error in the 
parallax and the above error in the bolometric flux, 
the total uncertainty in the luminosity is $\pm 850$~\LSOL.  

\begin{figure}
\resizebox{\hsize}{!}{\rotatebox{0}{\includegraphics{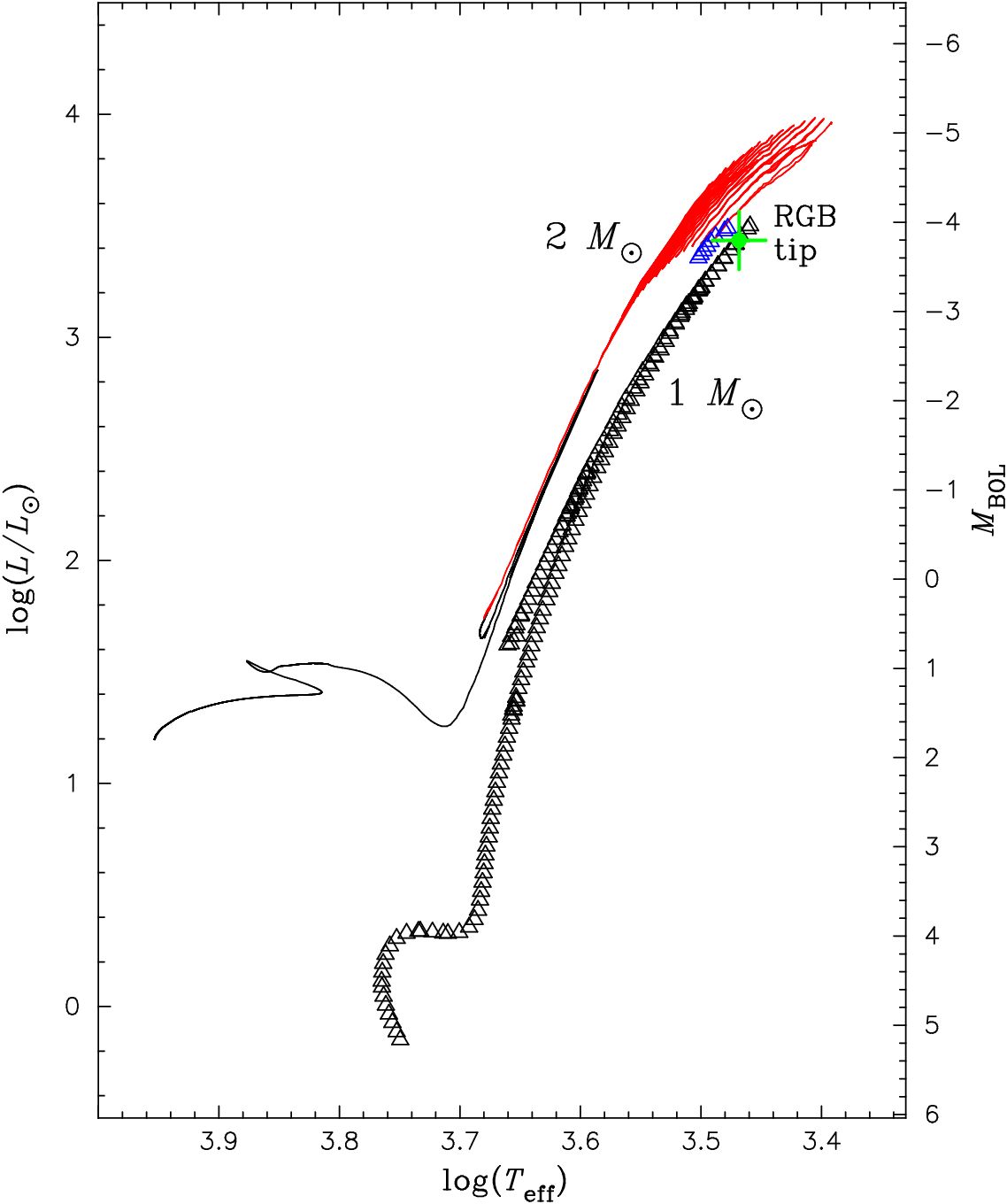}}}
\caption{Observationally derived position of BK~Vir on the H-R diagram 
(green dot with the error bars).  
The solid line and open triangles represent the evolutionary tracks 
for a 1 and 2~\MSOL\ star from Herwig (\cite{herwig05}) and Bertelli et al. 
(\cite{bertelli08}), respectively.  
The AGB stage (early AGB and thermally-pulsing AGB) is marked in blue and 
red for the evolutionary tracks for 1~\MSOL\ and 2~\MSOL, respectively. 
}
\label{hr_diagram}
\end{figure}

To estimate the stellar mass, we compared the position of BK~Vir on the 
H-R diagram with theoretical evolutionary tracks.
Figure~\ref{hr_diagram} shows evolutionary tracks for 1 and 2~\MSOL\ stars 
(Herwig \cite{herwig05}\footnote{http://astrowww.phys.uvic.ca/\textasciitilde
  fherwig} and Bertelli et al. \cite{bertelli08}, respectively), 
together with the observationally derived position of BK~Vir.  
The figure suggests that the mass of BK~Vir is close to 1~\MSOL. 
The adoption of a stellar mass of 1~\MSOL\ translates into a surface 
gravity of $\log \varg = -0.17$ with a stellar radius of 201~\RSOL\ (0.94~AU). 

Whereas BK~Vir is classified as an AGB star in most of the literature, 
Fig.~\ref{hr_diagram} shows that the luminosity of BK~Vir is just at the tip 
of the RGB.  
Moreover, Lebzelter \& Hron (\cite{lebzelter99}) report negative detection 
of $^{99}$Tc, which would be strong evidence of the third dredge-up in the 
AGB.  Therefore, we cannot conclude whether BK~Vir is an RGB star or an 
early-AGB star that has not yet experienced the third dredge-up.  
In either case, however, the surface chemical composition is 
expected to be changed by the first dredge-up in the RGB phase, which 
mixes the CN-cycled material to the stellar surface.  
This makes the MARCS models with the ``moderately CN-cycled'' chemical 
composition (with [Fe/H] = 0.0) suitable for BK~Vir.  

No analysis of the micro-turbulent velocity in BK~Vir is available in the 
literature.  However, the high-resolution spectroscopic analysis of M7--8 
giants with $\TEFF \la 3000$~K by Tsuji (\cite{tsuji08}) shows micro-turbulent 
velocities of 3--4~\KMS.  We adopt these values for BK~Vir as well.

\begin{figure}
\resizebox{\hsize}{!}{\rotatebox{0}{\includegraphics{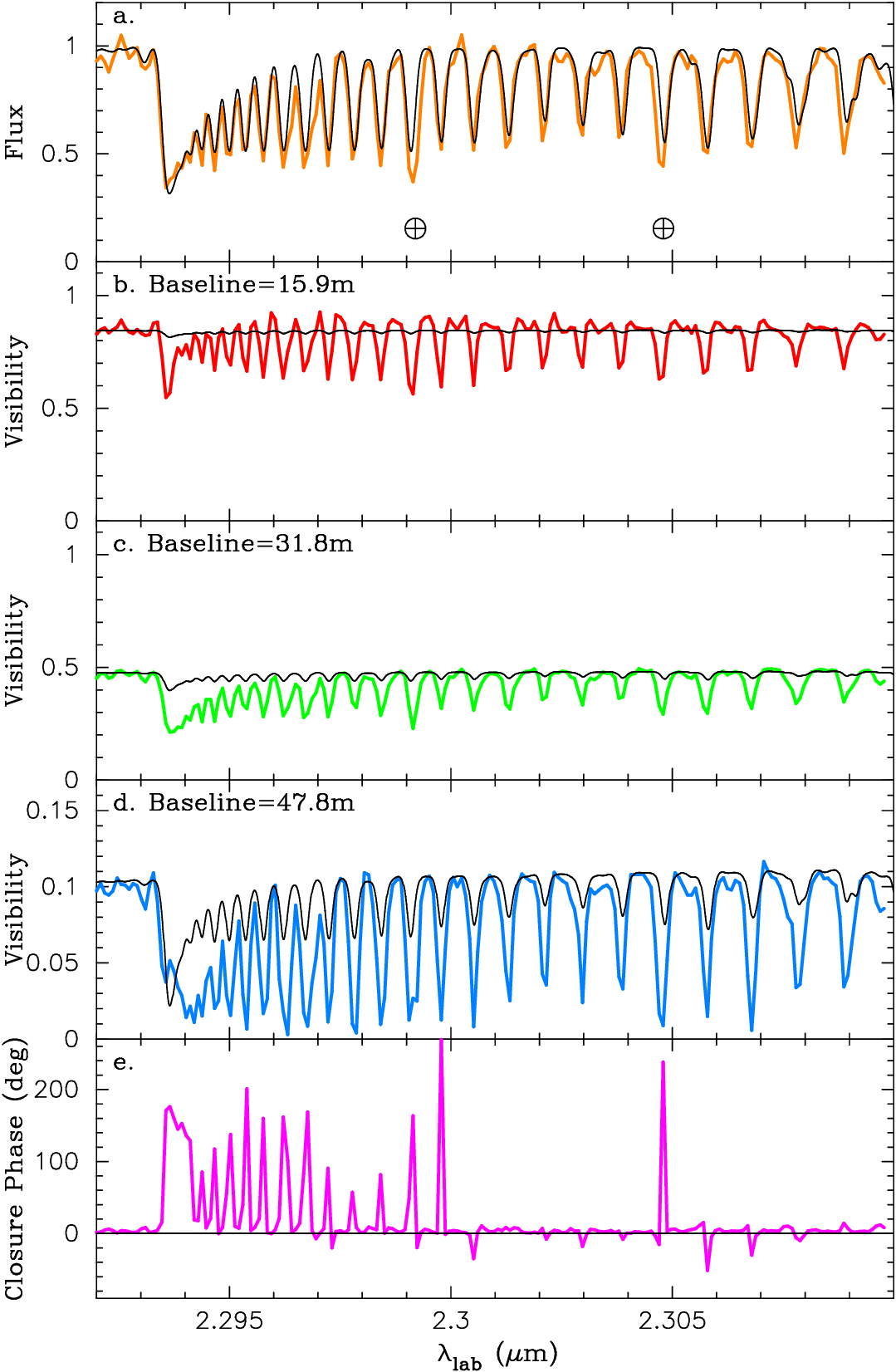}}}
\caption{
Comparison between the observed interferometric data (data set \#3) 
and the MARCS model. 
In all panels, the black thin lines represent the predictions of the MARCS 
model with the parameters most appropriate for BK~Vir (see
Sect.~\ref{subsect_marcs}).  The thick colored lines represent the observed 
data. 
{\bf a}) Spectrum.  Residuals of the strong telluric lines are marked 
by $\oplus$. 
{\bf b})--{\bf d}) Visibilities on the 15.9, 31.8, and 47.8~m baselines, 
respectively. 
{\bf e}) Closure phase. 
}
\label{marcs3000}
\end{figure}

\begin{figure}
\resizebox{\hsize}{!}{\rotatebox{0}{\includegraphics{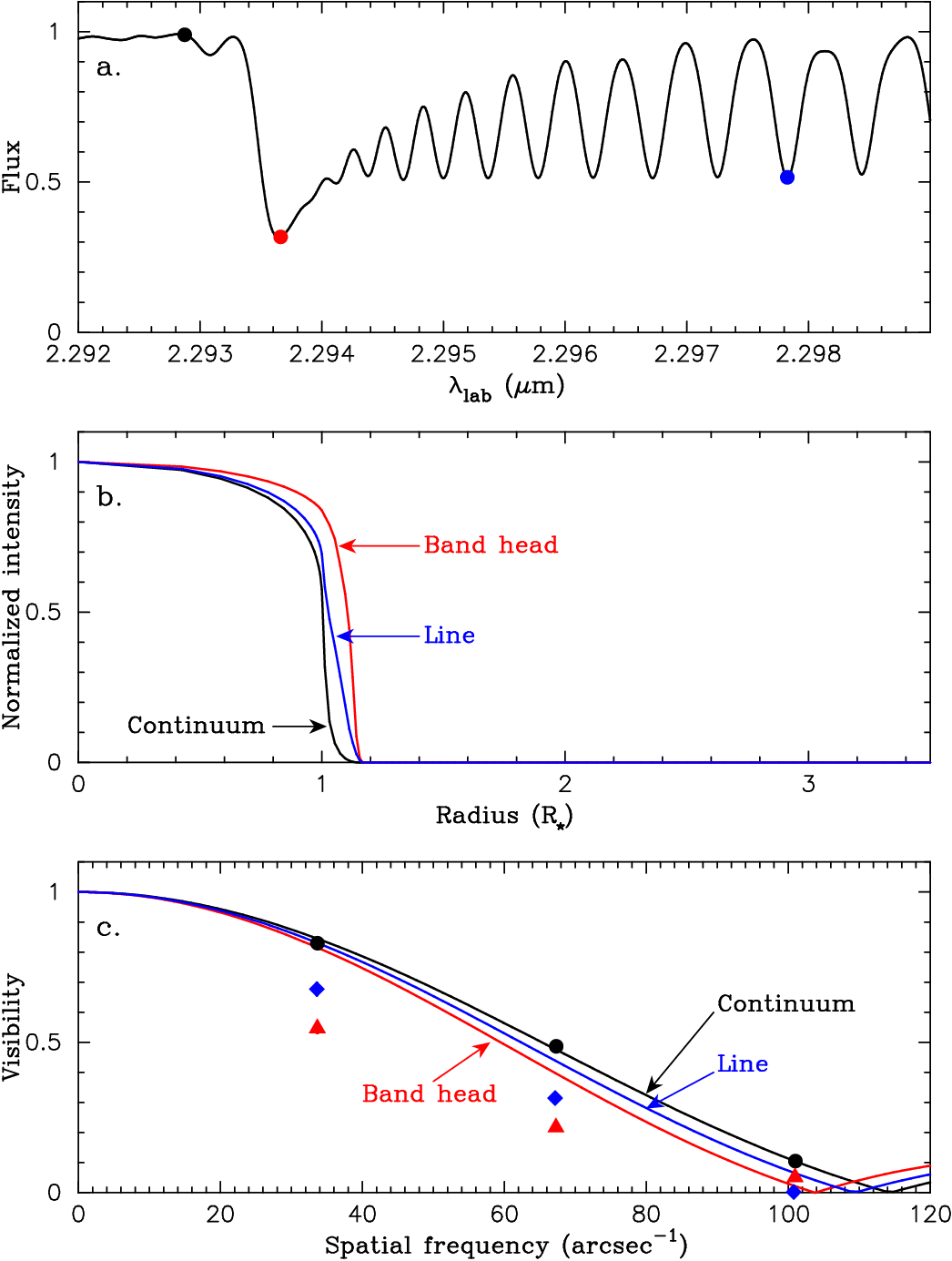}}}
\caption{
MARCS model for BK~Vir.  
{\bf a}) Model spectrum near the CO band head.  The black, red, and blue 
dots represent the wavelengths of the intensity profiles and visibilities 
shown in the panels {\bf b} and {\bf c}.  
{\bf b}) Model intensity profiles for the continuum (black), CO band 
head (red) and an isolated CO line (blue). 
{\bf c}) Model visibilities for the continuum, CO band head, and an 
isolated CO line are shown by the black, red, and blue solid lines, 
respectively.  
The filled symbols represent the observed visibilities 
(data set \#3) at the corresponding wavelengths. 
}
\label{clvvisplot_marcs3000}
\end{figure}

\subsection{Comparison with the MARCS photospheric model}
\label{subsect_marcs}

The MARCS model with the parameters 
as close to the derived values as possible is characterized by 
\TEFF\ = 3000~K, $\log \varg$ = 0.0, $M_{\star}$ = 1~\MSOL, \VMICRO\ = 2~\KMS\ 
(models with \VMICRO\ = 3--4~\KMS\ are not available for this combination 
of \TEFF, $\log \varg$, $M_{\star}$, and chemical composition, and the model 
with \VMICRO\ = 2~\KMS\ is the closest one).  
Using the temperature and pressure distributions of the downloaded MARCS 
model, we computed the monochromatic intensity profile and spectrum over 
the observed wavelength range using the CO line list of Goorvitch 
(\cite{goorvitch94}), as described in Ohnaka et al. (\cite{ohnaka06}).  
The monochromatic visibility was calculated by the Fourier transform 
of the monochromatic intensity profile. 
The intensity profile, visibility, and spectrum were then convolved with 
the AMBER's spectral resolution.  
The angular scale of the model was determined so that the uniform-disk 
diameter derived from the model visibilities in the continuum matches 
the adopted value of 10.73~mas.  

Figure~\ref{marcs3000} shows a comparison between the observed data 
(data set \#3) and the predictions 
by the MARCS photospheric model.  While the observed CO spectrum is well 
reproduced, the predicted visibilities in the CO lines are too 
high compared to the observed data.  
We also used other available MARCS models with slightly different stellar 
parameters (\TEFF\ = 2700--3100~K, \LOGG\ = $-0.5$ -- $+0.5$, $M_{\star}$ = 
1--2~\MSOL, and \VMICRO\ = 2--5~\KMS), but 
they cannot reproduce the observed visibilities in the CO lines, either. 
Figure~\ref{clvvisplot_marcs3000} shows the intensity profiles and 
visibilities at three representative wavelengths in the continuum, 
at the CO band head, and in an isolated line.  
The figure illustrates that the photosphere extends to 1.17~\RSTAR\ 
in the CO band head and in the isolated line, 
but this geometrical extension is insufficient to 
explain the observed visibilities (Fig.~\ref{clvvisplot_marcs3000}c).  
The too high visibilities in the CO lines predicted by the MARCS model 
means the actual CO-line-forming layer is much more extended than 
the model predicts or there is an additional component 
contributing to the CO lines above the photosphere modeled by MARCS. 
Therefore, our AMBER observations have spatially resolved the MOLsphere 
in a normal M giant for the first time in the individual CO lines.  
The same trend was reported for the normal M6-7 giant RS~Cap 
by Mart\'{i}-Vidal et al. (\cite{marti-vidal11}) based on AMBER observations 
of the CO bands with a lower spectral resolution of 1500.  
These studies illustrate that testing photospheric models only with 
the spectra of strong molecular or atomic features can be misleading.  

\begin{table*}
\begin{center}
\caption {Parameters of the MARCS+MOLsphere model for BK~Vir}
\vspace*{-2mm}

\begin{tabular}{l l l}\hline
Parameter &  Searched range &  Solution \\
\hline
CO column density of the inner layer: $N_{\rm inner}$ (\PERSQCM) & 
$10^{21}, 2\times10^{21}, 5\times10^{21}, 10^{22}, ..., 
10^{23}$ & (1--2)$\times10^{22}$ \\
CO column density of the outer layer: $N_{\rm outer}$ (\PERSQCM) & 
$5\times10^{18}, 10^{19}, 2\times10^{19}, 5\times10^{19}, 
..., 5\times10^{20}$ & $10^{19}$--$10^{20}$ \\
Temperature of the inner layer: $T_{\rm inner}$ (K) &
1800 ... 2300 ($\Delta T_{\rm inner}$ = 100~K) & 1900--2100\\
Temperature of the outer layer: $T_{\rm outer}$ (K) &
1200 ... 2300 ($\Delta T_{\rm outer}$ = 100~K) & 1500--2100\\
Radius the inner layer: $R_{\rm inner}$ (\RSTAR) &
1.2, 1.25, 1.3 & 1.2--1.25\\
Radius the outer layer: $R_{\rm outer}$ (\RSTAR) &
1.5, ..., 3.5 ($\Delta R_{\rm outer}$ = 0.5~\RSTAR) & 2.5--3.0\\
\hline
\label{table_param}
\vspace*{-7mm}

\end{tabular}
\end{center}
\end{table*}

\subsection{Modeling of the MOLsphere}
\label{subsect_wme}

To characterize the physical properties of the MOLsphere, 
we interpreted the AMBER data with a model in which one or 
two extended CO layers with constant temperatures and densities 
are added above the MARCS photospheric model.  
This model is the same as that described in Ohnaka (\cite{ohnaka04}), 
but with the following modification: 
each layer is assumed to be geometrically thin with a thickness of 
0.1~\RSTAR.  The inner radius of each layer is used as the parameter to 
represent the size of the layer.  
The radius, temperature, 
and column density of each layer are changed as free parameters to 
fit the observed spectrum and visibilities simultaneously.  
We tentatively adopted a microturbulent velocity of 4~\KMS\ for the 
CO layers based on the spectroscopic analysis of Tsuji (\cite{tsuji08}) 
for M7--8 giants.  
We first attempted to explain the observed data with models with one 
additional CO layer but could not reproduce the spectrum and visibilities 
simultaneously.  
Therefore, we attempted to explain the data with the models with two CO 
layers.  
Table~\ref{table_param} gives the range of the parameters 
searched in our modeling.  Note that the radius of the inner layer 
was set to be equal to or larger than 1.2~\RSTAR\ to avoid an overlap 
with the MARCS photospheric model, which extends to 1.17~\RSTAR. 

\begin{figure}
\resizebox{\hsize}{!}{\rotatebox{0}{\includegraphics{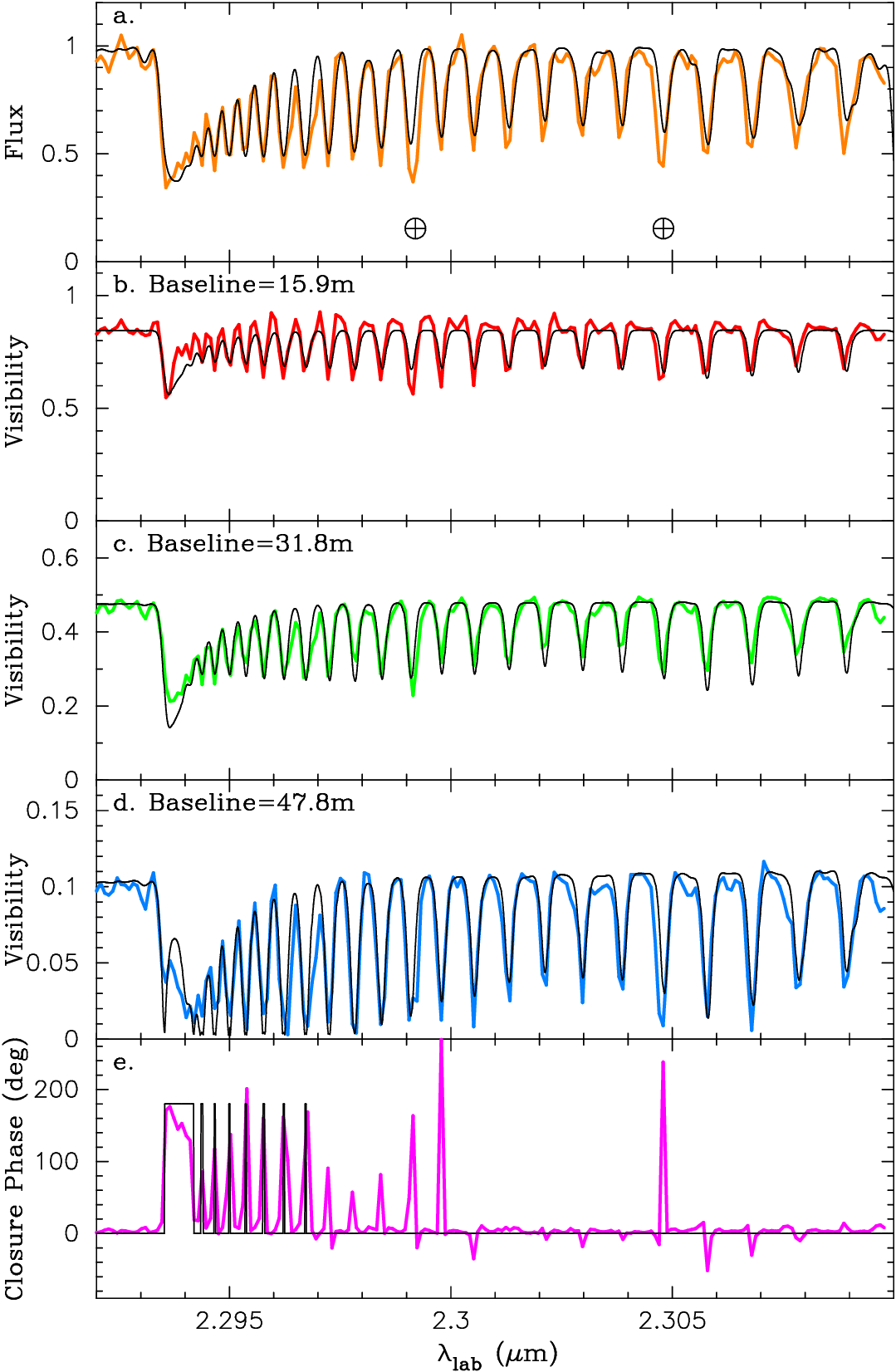}}}
\caption{
Comparison between the observed interferometric data and the MARCS+MOLsphere 
model shown in the same manner as Fig.~\ref{marcs3000}.  
}
\label{marcs3000_wme_final}
\end{figure}

\begin{figure}
\resizebox{\hsize}{!}{\rotatebox{0}{\includegraphics{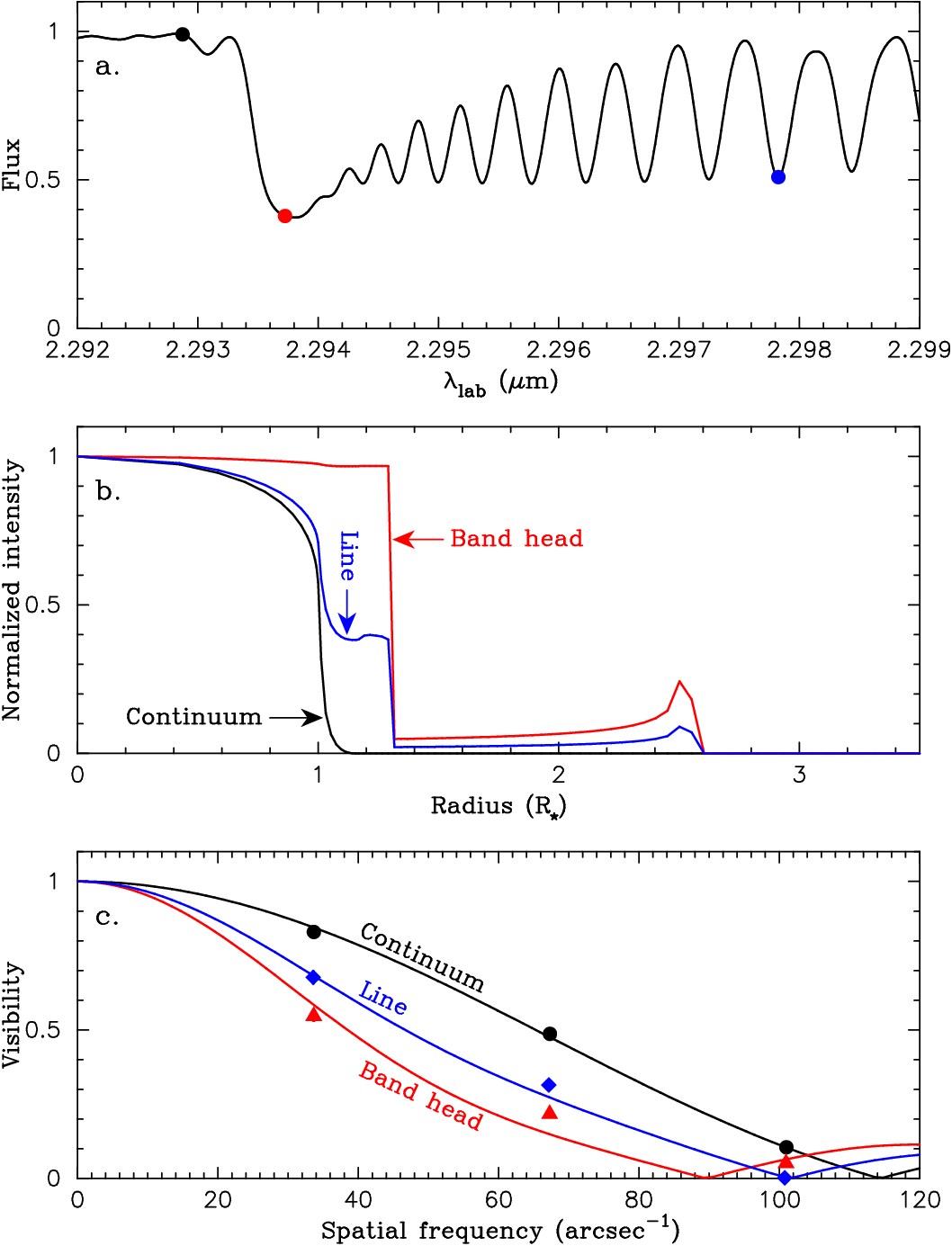}}}
\caption{
MARCS+MOLsphere model for BK~Vir shown in the same manner as 
Fig.~\ref{clvvisplot_marcs3000}. 
}
\label{clvvisplot}
\end{figure}

Figure~\ref{marcs3000_wme_final} shows a comparison between the observed data 
(data set \#3) and the best-fit model.  
The model intensity profiles and visibilities at three representative 
wavelengths (continuum, CO band head, and isolated CO line) are plotted 
in Fig.~\ref{clvvisplot}.  
This model is characterized by the inner CO layer at 1.2~\RSTAR\ with 
2000~K and a CO column density of $2\times10^{22}$~\PERSQCM\ and the outer CO 
layer at 2.5~\RSTAR\ with 1900~K and a column density of $2\times10^{19}$~\PERSQCM. 
The column densities of the inner and outer CO layer correspond to 
16\% and 0.016\% of the CO column density in the MARCS photospheric model. 
We found the range of the radius, temperature and CO column density 
of the inner layer to be 1.2--1.25~\RSTAR, 1900--2100~K, and 
(1--2)$\times10^{22}$~\PERSQCM, respectively.  
The derived ranges for the radius, temperature, and CO column density 
of the outer layer are 2.5--3.0~\RSTAR, 1500--2100~K, and 
$10^{19}$--$10^{20}$~\PERSQCM, respectively.  
As shown in Fig.~\ref{marcs3000_wme_final}, 
the observed visibilities and spectrum are well reproduced.  
The model also predicts the CP to jump in the CO lines, in particular at the 
CO band head, as seen in the AMBER data.  
This can be explained as follows.  Figure~\ref{clvvisplot}b shows that 
the inner CO layer, which is optically thick, makes the star appear larger 
in the CO band head than in the continuum by 30\%.  Then the 
longest baseline samples the second visibility lobe in the CO band 
head, as shown in Fig.~\ref{clvvisplot}c (red line and red triangle at 
a spatial frequency of $\sim$100~arcsec$^{-1}$).  
The phase in the second visibility lobe is $\pi$, while the phases 
on the shortest and middle baselines in the first visibility lobe are zero.  
This results in a predicted CP of $\pi$ for the CO band head and 
leads to the jumps in the CP as observed.  
Because the model is spherically symmetric, 
the observed non-zero/non-$\pi$ CPs and non-zero DPs are not 
reproduced\footnote{The model DPs for the shortest and middle baselines 
are zero, and the model DP for the longest baseline is the same as the 
model CP with its sign flipped.}.      
However, the overall agreement in the spectrum, visibilities, and CP 
suggests that our model represents the approximate picture of the outer 
atmosphere of BK~Vir.

It is impossible to uniquely constrain the detected inhomogeneous
structure from our observations.  However, we estimated the flux contribution 
of asymmetric features from the observed CPs as follows.  
We added a Gaussian-shaped spot to the best-fit model image in the CO lines 
and computed the visibility and CP at the observed baselines.  
We changed the size and intensity of the Gaussian-shaped spot so that 
the observed non-zero/non-$\pi$ CPs (40--100\degr) can be 
reproduced without affecting the fit to the observed visibilities 
significantly.  
This experiment suggests that the flux contribution of the asymmetric  
feature is 3--5\% of the total flux.  
These values agree with those estimated for other AGB stars.  
Ragland et al. (\cite{ragland06}) derived a flux 
contribution of 3\% from a geometrical modeling (uniform disk + 
unresolved spot) of the asymmetric features detected in a sample of AGB 
stars in the $H$-broadband.  
Wittkowski et al. (\cite{wittkowski11}) also suggest a flux contribution 
of up to $\sim$3\% in the 2~\micron\ \HOH\ band for the Mira star R~Cnc.

The observed spectrum is nearly equally well reproduced by the 
MARCS-only model and MARCS+MOLsphere model, despite the clear 
disagreement between the observed visibilities and MARCS-only model.  
The reason is that the additional absorption due to the CO layers above the 
MARCS model is filled in by the extended emission of the same 
CO layers in the spatially unresolved spectrum.  Therefore, good agreement 
between observation and model for spatially unresolved spectra of strong 
molecular features can be misleading, because these features can include 
the contribution from the extended outer atmosphere.  

\section{Discussion}
\label{sect_discuss}

\begin{figure}
\resizebox{\hsize}{!}{\rotatebox{-90}{\includegraphics{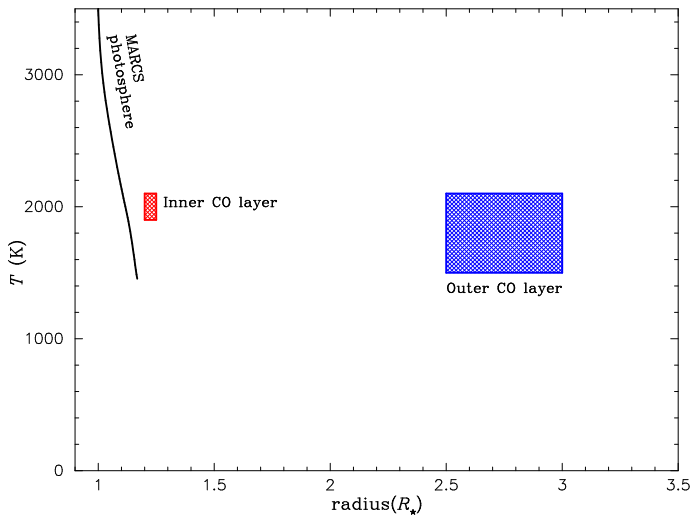}}}
\caption{
Temperature distribution as a function of radius.  The black solid line 
represents the temperature profile of the MARCS model used for BK~Vir. 
The red and blue rectangles represent the ranges of the inner and outer 
CO layers derived from our modeling, respectively.  
}
\label{plotTemp}
\end{figure}

Figure~\ref{plotTemp} shows the temperature distribution of the MARCS 
photospheric model as a function of radius, together with the location and 
the temperature of 
the extended CO layers derived from our modeling.  The temperature of the 
inner CO layer is noticeably higher than that of the uppermost layer of the 
MARCS photospheric model, although the height is approximately the same. 
The temperature of the outer CO layer is also 
higher than or equal to that of the top of the photosphere.  
This suggests a temperature inversion in the MOLsphere of BK~Vir.  

Given that our spectroscopic calibration method without a usual spectroscopic 
standard star gives the reliable CO line spectra of BK~Vir as described in 
Sect.~\ref{sect_obs}, the uncertainty in the spectra are unlikely to affect 
the result of our modeling.  

We checked whether this is merely a result of approximating the 
temperature distribution in the outer atmosphere by two discrete layers.  
We set the temperatures of the CO layers to be below 1500~K, which is the 
temperature of the uppermost layer of the MARCS photospheric model, 
and attempted to reproduce the observed data.  However, it turned out that 
such models cannot provide a reasonable fit to the observed spectrum and 
visibilities simultaneously.  
Therefore, 
the temperature inversion suggested in Fig.~\ref{plotTemp} is unlikely 
to be a result of the simplification we adopted for the structure of the 
outer atmosphere. 

We also checked whether or not the assumption of local thermodynamical 
equilibrium (LTE) used in our model is valid for the CO layers. 
As described in Ohnaka (\cite{ohnaka04}), 
we compared the collisional and radiative 
de-excitation rates.  The collisional de-excitation rate $C_{\rm ul}$ is 
estimated as $N \, \sigma_{\rm ul} \,\varv_{\rm rel}$, where $N$ is the number 
density of the primary collision partner, $\sigma_{\rm ul}$ is the cross 
section, which is approximated with the geometrical cross section, and 
$\varv_{\rm rel}$ is the relative velocity between the collision partner 
and CO molecules.  The CO column density of $10^{21}$~\PERSQCM\ derived 
for the inner layer and the layer's geometrical thickness of 0.1~\RSTAR\ 
translates into a CO number density of $7 \times 10^{9}$~\PERCBCM.  
In chemical equilibrium at the temperature of 2000~K derived for the inner 
layer, number densities of H atoms and H$_2$ molecules of $1.2\times10^{13}$ 
and $1.4\times10^{13}$~\PERCBCM\ are necessary to obtain the 
above CO number density.  We take both H atoms and H$_2$ molecules as the 
primary collision partner of CO molecules, which leads to $N = 2.6 \times 
10^{13}$~\PERCBCM.  With a geometrical cross section $\sigma_{\rm ul}$ of 
$10^{-15}$~\PERSQCM\ and a relative velocity $\varv_{\rm rel}$ of 5~\KMS 
assumed, we obtain $C_{\rm ul} = 1.3\times10^{4}$~s$^{-1}$ for the inner 
CO layer.  
In the case of the outer CO layer, 
the lower limit of the CO column density ($10^{19}$~\PERSQCM) translates into 
a CO number density of $7.1\times10^{6}$~\PERCBCM.  
For the temperatures of 1500--2100~K derived for the outer layer, 
the number densities of H atoms and H$_2$ molecules are $5.0\times10^{9}$ 
and $1.6\times10^{10}$~\PERCBCM\ (1500~K) and $3.8\times10^{10}$ and 
$4.2\times10^{7}$~\PERCBCM\ (2100K), respectively.  This leads to 
$N$ (H + H$_2$) = (2--4)$\times 10^{10}$~\PERCBCM.  With the above $\sigma_{\rm ul}$ 
and $\varv_{\rm rel}$, we obtain $C_{\rm ul} = 10$~s$^{-1}$ for the outer CO 
layer.  
These collisional de-excitation rates are much 
larger than radiative de-excitation rates $A_{\rm ul}$ of 0.6--1~s$^{-1}$ 
found for the CO lines analyzed in the present work, which means the 
validity of LTE for our modeling.  
Therefore, it is unlikely that the high temperatures derived for the 
CO layers are just a result of non-LTE effects.  
 
The temperature inversion implies the operation of some heating mechanism 
in the outer atmosphere within $\sim$3~\RSTAR.  
While the outwardly propagating shocks generated by the stellar pulsation 
can produce temperature jumps in Mira stars (e.g., Richter et al. 
\cite{richter03}), it is not clear whether this 
is the case for normal M giants, where the variability amplitudes are 
much smaller.  Despite the remarkable difference in the variability 
amplitude, BK~Vir shows the temperature inversion in the molecular 
component of the outer atmosphere.  
This implies that the stellar pulsation may not be the cause of the heating.  

Some kind of heating is also required to explain the chromosphere common 
among M giants, 
although no UV or H$\alpha$ observations of BK~Vir are available in the 
literature. 
However, 
the (electron) temperature of the chromosphere inferred from the UV emission 
lines in M giants 
is much higher ($\sim$6000---10000~K, Eaton \& Johnson \cite{eaton88}; 
Luttermoser et al. \cite{luttermoser94}) 
than the 1900--2100~K derived for the inner CO layer.  
Therefore, although the inner CO layer is warmer than the uppermost layer of 
the photosphere, the CO gas in the extended outer atmosphere represents a 
distinct component other than the chromosphere.  

Still, the process responsible for the chromospheric heating may lead to the 
formation of both of the hot plasma and warm molecular gas in the outer 
atmosphere.  
For example, the magnetohydrodynamical (MHD) simulation of stellar winds 
in red giants by Suzuki (\cite{suzuki07}) shows the formation of transient 
hot gas bubbles embedded in cool gas outflows, 
although the simulation is still limited 
to stars with relatively high effective temperatures ($\ga$3900~K) and 
much less luminous stars ($\la$1000~\LSOL) compared to BK~Vir.  
However, as 
Airapetian et al. (\cite{airapetian10}) point out,  this simulation 
assumes 
fully-ionized plasma, although the gas around K--M giants is only weakly 
ionized.  The MHD simulation of Airapetian et al. (\cite{airapetian10}) 
reproduces the observed terminal wind velocities and mass-loss rates 
of K giants, but the agreement between the modeled and observed radial 
velocity fields is not yet satisfactory (comparison between the model of 
Airapetian et al. \cite{airapetian10} and observed data summarized in 
Crowley et al. \cite{crowley09} is shown in Harper \cite{harper11}).  
Since the simulation of Airapetian et al. (\cite{airapetian10}) 
assumes a nearly iso-thermal chromosphere, it 
does not include the cool, molecular component.  
The extension of these MHD simulations to cooler and more luminous 
stars including the modeling of thermal structures 
would be useful for understanding the origin of the MOLsphere 
in normal red giants.  

The presence of the dense CO layers above the photosphere suggests that there 
can also be other molecular species in the MOLsphere.  For example, TiO is one 
of such molecules, and the contribution from the MOLsphere can significantly 
affect the strength of the TiO bands observed in the optical.  
Hron et al. (\cite{hron10}) show that the MOLsphere models for the red 
supergiant Betelgeuse available in the 
literature predict the TiO bands to be too 
strong compared to the observed ones.  This is because of the strong 
contribution from the TiO molecules in the MOLsphere.  
As possible reasons, they suggest the clumpy or patchy nature of the MOLsphere 
with a small filling factor and/or the crude nature of the current MOLsphere 
models.  In addition, we suggest that the TiO bands may form by scattering 
in the MOLsphere.  The line formation by scattering for electronic transitions 
(such as the TiO bands) is suggested by Hinkle \& Lambert (\cite{hinkle75}).  
The construction of the MOLsphere model with this process taken into account 
is beyond the scope of this paper, but should be pursued to obtain a 
more comprehensive picture of the MOLsphere.

\section{Concluding remarks}
\label{sect_concl}

We have spatially resolved the normal M7 giant BK~Vir in the individual 
CO first overtone lines with VLTI/AMBER.  
The uniform-disk diameter shows clear increases by 12--31\% in the 
CO lines compared to the continuum.  We detected non-zero/non-$\pi$ 
CPs and non-zero DPs in the CO lines, which reveal an asymmetry in 
the upper photosphere and outer atmosphere where the CO lines form.  
The visibilities, CPs, and DPs observed in the individual CO lines are 
symmetric with respect to the line center.  This is in marked contrast 
with the red supergiant Betelgeuse, in which the vigorous inhomogeneous 
gas motions manifest themselves as the asymmetry in the visibilities 
in the blue and red wing of the lines.  The absence of such asymmetry 
indicates that the amplitude of the inhomogeneous velocity field in 
BK~Vir is much smaller ($\la$5~\KMS) than in Betelgeuse. 

While the MARCS photospheric model can well reproduce the observed CO line 
spectrum of BK~Vir, the angular size in the CO lines 
predicted by the model is too small.  
This indicates the presence of more extended 
CO layers than predicted by the photospheric model.  Our model with two 
CO layers added above the MARCS model can explain the observed 
spectrum, visibilities, and CPs reasonably well.  The derived temperatures 
of the two extended CO layers, 1900--2100~K and 1500--2100~K, 
are higher than or equal to 
that of the uppermost layer of the photosphere.  This result suggests the 
operation of some heating mechanism in the outer atmosphere within 
$\sim$3~\RSTAR.  

The present work illustrates that AMBER observations with high-spectral 
resolution are effective for constraining the physical properties of the 
outer atmosphere.  Given that normal red giants with small variability 
amplitudes are not yet intensively 
studied by IR spectro-interferometry, it is indispensable to carry out a 
systematic survey of a sample of these stars with ranges of 
temperatures and luminosities.  This will shed new light on the mass-loss 
mechanism not only in normal red giants but also in Mira stars.

\begin{acknowledgement}
We thank the ESO VLTI team for supporting our AMBER observations.  
NSO/Kitt Peak FTS data on the Earth's telluric features 
were produced by NSF/NOAO.
This publication makes use of data products from the Two Micron All Sky
Survey, which is a joint project of the University of Massachusetts and the
Infrared Processing and Analysis Center/California Institute of Technology,
funded by the National Aeronautics and Space Administration and the National
Science Foundation.
\end{acknowledgement}


\begin{thebibliography}{}

\bibitem[2010]{airapetian10}
Airapetian, V. S., Carpenter, K. G., \& Ofman, L.\ 2010, ApJ, 723, 1210

\bibitem[1997]{bedding97}
Bedding, T., Zijlstra, A. A., von der L\"uhe, O., et al.\ 1997, MNRAS, 286, 
957

\bibitem[2008]{bertelli08}
Bertelli, G., Girardi, L., Marigo, P., \& Nasi, E.\ 2008, A\&A, 484, 815

\bibitem[2009]{crowley09}
Crowley, C., Espey, B. R., Harper, G. M., \& Roche, J.\ 2009, 
Proceedings of the 15th Cambridge Workshop on Cool Stars, Stellar Systems and 
the Sun, AIP Conference Proceedings, Vol. 1094, p.267

\bibitem[2010]{debeck10}
De Beck, E., Decin, L., de Koter, A., et al.\ 2010, A\&A, 523, 18

\bibitem[1998]{dyck98}
Dyck, H. M., van Belle, G. T., \& Thompson, R. R.\ 1998, AJ, 116, 981

\bibitem[1995]{eaton95}
Eaton, J. A.\ 1995, AJ, 109, 1797

\bibitem[1988]{eaton88}
Eaton, J. A., \& Johnson, H. R.\ 1988, ApJ, 325, 355

\bibitem[2005]{famaey05}
Famaey B., Jorissen A., Luri X., et al.\ 2005, A\&A, 430, 165

\bibitem[1992]{fouque92}
Fouque, P., Le Bertre, T., Epchtein, N., Guglielmo, F., \& Kerschbaum, F.\ 
1992, A\&AS, 93, 151

\bibitem[2006]{gontcharov06}
Gontscharov, G. A.\ 2006, Astron. Lett., 32, 759

\bibitem[2003]{gonzalez_delgado03}
Gonz\'alez Delgado, D., Olofsson, H., Kerschbaum, F., et al.\ 2003, A\&A, 
411, 123

\bibitem[1994]{goorvitch94}
Goorvitch, D.\ 1994, ApJS, 95, 535

\bibitem[2008]{gustafsson08}
Gustafsson, B., Edvardsson, B., Eriksson, K., et al.\ 2008, A\&A, 486, 951

\bibitem[2011]{harper11}
Harper, G.\ 2011, Presentation at ``From Atoms to Stars: 
the impact of Spectroscopy on Astrophysics'', \\
http://media.atomstars.org/presentations/harper.pdf

\bibitem[2005]{herwig05}
Herwig, F.\ 2005, ARA\&A, 43, 435

\bibitem[1975]{hinkle75}
Hinkle, K. H., \& Lambert, D. L.\ 1975, MNRAS, 170, 447

\bibitem[2010]{hron10}
Hron, J., Aringer, B., \& Paladini, C.\ 2010, Poster presentation at 
``The Origin and Fate of the Sun: Evolution of Solar-mass Stars Observed with 
High Angular Resolution'', \\
http://www.eso.org/sci/meetings/2010/stars2010/Presentations\\
/P-hron\_molsphere-poster.pdf

\bibitem[2011]{isabel_perez_martinez11}
Isabel P\'erez Mart\'inez, M., Schr\"oder, K.-P., \& Cuntz, M.\ 2011, MNRAS, 
414, 418

\bibitem[2007]{ita07}
Ita, Y., Tanab\'e, T., Matsunaga, N., et al.\ 2007, PASJ, 59, S437

\bibitem[1997]{karovska97}
Karovska, M., Hack, W., Raymond, J., \& Guinan, E.\ 1997, ApJ, 482, L175

\bibitem[1994]{kerschbaum94}
Kerschbaum, F., Hron, J.\ 1994, A\&AS, 106, 397

\bibitem[1995]{lebzelter95}
Lebzelter, T., Kerschbaum, F., \& Hron, J.\ 1995, A\&A, 298, 159

\bibitem[1999]{lebzelter99}
Lebzelter, T., \& Hron, J.\ 1999, A\&A, 351, 533

\bibitem[1995]{luck95}
Luck, R. E., \& Wepfer, G. G.\ 1995, AJ, 110, 2425

\bibitem[1994]{luttermoser94}
Luttermoser, D. G., Johnson, H. R., \& Eaton, J. A.\ 1994, ApJ, 422, 351

\bibitem[2010]{lyubimkov10}
Lyubimkov, L. S., Lambert, D. L., Rostopchin, S. I., Rachkovskaya, T. M., 
\& Poklad, D. B.\ 2010, MNRAS, 402, 1369

\bibitem[2011]{marti-vidal11}
Mart\'i-Vidal, I., Marcaide, J. M., Quirrenbach, A., et al.\ 2011, A\&A, 529, 
A115

\bibitem[2000]{mcmurry00}
McMurry, A. D., \& Jordan, C.\ 2000, MNRAS, 313, 423

\bibitem[2004]{mondal04}
Mondal, S., \& Chandrasekhar, T.\ 2004, MNRAS, 348, 1332

\bibitem[1969]{neugebauer69}
Neugebauer, G., \& Leighton, R. B.\ 1969, Two-micron sky survey 
(NASA SP-3047, Washington)

\bibitem[2004]{ohnaka04}
Ohnaka, K.\ 2004, A\&A, 424, 1011

\bibitem[2005]{ohnaka05}
Ohnaka, K., Bergeat, J., Driebe, T., et al.\ 2005, A\&A, 429, 1057

\bibitem[2006]{ohnaka06}
Ohnaka, K., Scholz, M., \& Wood, P.\ 2006, A\&A, 446, 1119

\bibitem[2009]{ohnaka09}
Ohnaka, K., Hofmann, K.-H., Benisty, M., et al.\ 2009, A\&A, 503, 183

\bibitem[2011]{ohnaka11}
Ohnaka, K., Weigelt, G., Millour, F., et al.\ 2011, A\&A, 529, A163

\bibitem[2007]{origlia07}
Origlia, L., Rood, R., Fabbri, S., et al.\ 2007, ApJ, 667, L85

\bibitem[2010]{origlia10}
Origlia, L., Rood, R., Fabbri, S., et al.\ 2007, ApJ, 718, 522

\bibitem[2003]{perrin03}
Perrin, G.\ 2003, A\&A, 400, 1173

\bibitem[1998]{perrin98}
Perrin, G., Coud\'e du Foresto, V., Ridgway, S. T., et al.\ 1998, A\&A, 
331, 619

\bibitem[2004]{perrin04}
Perrin, G., Ridgway, S. T., Coud\'e du Foresto, V., Mennesson, B., 
Traub, W. A., \& Lacasse, M. G.\ 2004, A\&A, 418, 675

\bibitem[2007]{petrov07}
Petrov, R. G., Malbet, F., Weigelt, G., et al.\ 2007, A\&A, 464, 1

\bibitem[2009]{pluzhnik09}
Pluzhnik, E., Ragland, S., le Coroller, E., et al.\ 2009, ApJ, 700, 114

\bibitem[2006]{ragland06}
Ragland, S., Traub, W. A., J.-P. Berger, et al.\ 2006, ApJ, 652, 650

\bibitem[2008]{ragland08}
Ragland, S., le Coroller, H., Pluzhnik, E., et al.\ 2008, ApJ, 679, 746

\bibitem[2005]{richichi05}
Richichi, A., \& Percheron, I.\ 2005, A\&A, 434, 1201

\bibitem[2003]{richter03}
Richter, He., Wood, P. R., Woitke, P., Bolick, U., \& Sedlmayr, E.\ 
2003, A\&A, 400, 319

\bibitem[2009]{samus09}
Samus, N. N.,  Durlevich, O. V., et al.\ 2009, General Catalogue of 
Variable Stars, GCVS database, Version 2011 Jan 
(Institute of Astronomy of Russian Academy of Sciences and Sternberg 
State Astronomical Institute of the Moscow State University, Moscow)

\bibitem[1998]{schlegel98}
Schlegel, D., Finkbeiner, D., \& Davis, M.\ 1998, ApJ, 500, 525

\bibitem[2006]{skrutskie06}
Skrutskie, M. F., Cutri, R. M., Stiening, R., et al.\ 2006, AJ, 131, 1163 
(The 2MASS All-Sky Catalog of Point Sources) 

\bibitem[2007]{suzuki07}
Suzuki, T. K.\ 2007, ApJ, 659, 1592

\bibitem[2009]{takami09}
Takami, H., Goto, M., Gaessler, W., et al.\ 2009, PASJ, 61, 623

\bibitem[2008]{takeda08}
Takeda, Y., Sato, B., \& Murata, D.\ 2008, PASJ, 60, 781

\bibitem[2007]{tatulli07}
Tatulli, E., Millour, F., Chelli, A., et al.\ 2007, A\&A, 464, 29

\bibitem[2001]{tsuji01}
Tsuji, T., 2001, A\&A, 376, L1

\bibitem[2008]{tsuji08}
Tsuji, T.\ 2008, A\&A, 489, 1271

\bibitem[1997]{tsuji97}
Tsuji, T., Ohnaka K., Aoki, W., \& Yamamura, I.\ 1997, A\&A, 320, L1

\bibitem[1999]{tuthill99}
Tuthill, P. G., Haniff, C. A., \& Baldwin, J. E.\ 1999, MNRAS, 306, 353

\bibitem[2007]{vanleeuwen07}
van Leeuwen, F.\ 2007, A\&A, 474, 653

\bibitem[1996]{wallace96}
Wallace, L., \& Hinkle, K. H.\ 1996, ApJS, 107, 312

\bibitem[1996]{weigelt96}
Weigelt, G., Balega, Y., Hofmann, K.-H., \& Scholz, M.\ 1996, A\&A, 316, L21

\bibitem[1994]{wiedemann94}
Wiedemann, G., Ayres, T. R., Jennings, D. E., \& Saar, S. H.\ 1994, ApJ, 423, 
806

\bibitem[2003]{winters03}
Winters, J. M., Le Bertre, T., Jeong, K. S., Nyman, L.-\AA., \& Epchtein, N.\ 
2003, A\&A, 409, 715

\bibitem[2007]{wittkowski07}
Wittkowski, M., Boboltz, D. A., Ohnaka, K., Driebe, T., \& Scholz, M.\ 2007, 
A\&A, 470, 191

\bibitem[2008]{wittkowski08}
Wittkowski, M., Boboltz, D. A., Driebe, T., Le Bouquin, J.-B., Millour, F., 
Ohnaka, K., \& Scholz, M.\ 2008, A\&A, 479, L21

\bibitem[2011]{wittkowski11}
Wittkowski, M., Boboltz, D. A., Ireland, M., et al.\ 2011, A\&A, 532, L7

\bibitem[2009]{woodruff09}
Woodruff, H. C., Ireland, M. J., Tuthill, P. G., et al.\ 2009, ApJ, 691, 1328

\bibitem[2004]{zacharias04}
Zacharias N., Monet D.G., Levine S.E., et al.\ 2004, 
BAAS, 36, 1418, 
Naval Observatory Merged Astrometric Dataset (NOMAD)

\end{thebibliography}
\end{document}